\documentclass[%
 reprint,
 amsmath,amssymb,
 aps,
]{revtex4-2}
\usepackage{booktabs}
\usepackage{multirow}
\usepackage{braket}
\usepackage{graphicx}% Include figure files
\usepackage{dcolumn}% Align table columns on decimal point
\usepackage{bm}% bold math
\usepackage{hyperref}% add hypertext capabilities
\begin{document}
\setlength{\parskip}{0pt}
\preprint{APS/123-QED}

\title{Variational Quantum Linear Solver with Dynamic Ansatz}% Force line breaks with \\

\author{Hrushikesh Patil}
\email{hrushikesh.patil@stonybrook.edu}
 \affiliation{%
 Department of Electrical and Computer Engineering, Stony Brook University, Stony Brook, NY 11794
}%

\author{Yulun Wang}%
 \email{yulun.wang@stonybrook.edu}
 \author{Predrag S. Krsti\'c}
\email{krsticps@gmail.com}
\affiliation{%
 Institute for Advanced Computational Science, Stony Brook University, NY 11794-5250
}%

\date{\today}% It is always \today, today,
    % but any date may be explicitly specified

\begin{abstract}
Variational quantum algorithms have found success in the NISQ era owing to their hybrid quantum-classical approach which mitigate the problems of noise in quantum computers. In our study we introduce the dynamic ansatz in the Variational Quantum Linear Solver for a system of linear algebraic equations. In this improved algorithm, the number of layers in the hardware efficient ansatz circuit is evolved, starting from a small and gradually increasing until convergence of the solution is reached. We demonstrate the algorithm advantage in comparison to the standard, static ansatz by utilizing fewer quantum resources and with a smaller quantum depth on average, in presence and absence of quantum noise, and in cases when the number of qubits or condition number of the system matrix are increased. The numbers of iterations and layers can be altered by a switching parameter. The performance of the algorithm in using quantum resources is quantified by a newly defined metric. 
\end{abstract}

\maketitle

%\tableofcontents

\section{\label{sec:level1}INTRODUCTION }

Solving a System of Linear Equations (SLE) is the fundamental problem in computation. Nearly 70\% of the scientific and engineering numerical problems reduces to solving SLE \cite{dahlquist_bjorck}, which can be symbolically expressed as finding unknown vector $\boldsymbol{x}$ of dimension $N$ from a system of equations $A\boldsymbol{x} = \boldsymbol{b}$ , where $A$ is a matrix of dimension $M \times N$, and $\boldsymbol{b}$ is a vector of dimension $M$. In this paper we will consider only square SLEs with $N=M$, since these appear most often in applications. In classical computation, solving of a SLE scales polynomially with $N$ \cite{SLE_complexity}. Well known quantum computing SLE algorithm HHL \cite{HHL} scales logarithmically with N. However, the application of HHL is severely limited in the Noisy Intermediate-Scale Quantum (NISQ) era due to its considerable quantum depth. The hybrid approaches, which combine quantum and classical computing \cite{Kandala2017, Cerezo2020variationalquantum, bravoprieto2020variational, huang2019nearterm, xu2019variational, bharti2021noisy}, result in a family of algorithms called Variational Hybrid Quantum Classical Algorithms (VHQCA). These have been successfully applied in computational chemistry (VQE) \cite{Kandala2017}), fidelity estimation \cite{Cerezo2020variationalquantum}, quantum machine learning \cite{Biamonte2017}, etc. Characteristics of a VHQCA is quantum ansatz of a small quantum depth, combined with a classical, post-processing optimization routines. Despite the promising results, VHQCAs have shown to be also susceptible to damaging effects of noise \cite{wang2021noiseinduced}, especially when the number of qubits increases. 

Bravo-Prieto et al. \cite{bravoprieto2020variational} proposed a VHQCA for solving a system of linear equations called Variational Quantum Linear Solver (VQLS). This algorithm uses a variational quantum circuit to evaluate the overlap between states $A\ket{x}$ and $\ket{b}$, where the state $\ket{x}$ is a trial wave function which approximates the solution of the SLE. This is encoded iteratively by a variational quantum circuit, ansatz, followed by the classical optimization of the variational parameters until convergence is reached. VQLS has polylogarithmic scaling with $N$ and was used to solve linear systems of size $N=1024$ with $n=log_2(N)=10$ qubits on a quantum computer \cite{bravoprieto2020variational}. This is a significant achievement as compared to the maximal system size 8 (i.e. with 3 qubits) achieved by the HHL algorithm \cite{expreal}. Similar VHQCA's for SLE have been proposed by Huang et al. \cite{huang2019nearterm} and Xu et al. \cite{xu2019variational}. However, since VQLS has polylogarithmic scaling (as $O((log_2(N))^{8.5}\kappa log(1/\epsilon))$ \cite{bravoprieto2020variational}, iterative classical methods like conjugate gradient (CG) which scale as $O(Ns\kappa log(1/\epsilon))$ \cite{HHL} are faster for small system sizes. VQLS is expected to show advantage over CG for very large system sizes, $N > 2 \times 10^{12}$ ($>$ 41 qubits) with sparsity $s = 0.5$, precision $\epsilon = 0.01$  and condition number $\kappa = 1$ . However, unlike VQLS the CG method is constrained to positive definite matrices. Hence, for classical iterative solvers like GMRES (Generalized Minimum Residual Method) which are more generalized which scale as $O(N^2)$\cite{Saad1986GMRESAG}, VQLS would perform better for matrix sizes larger than $2\times 10^5$ (corresponding to $>$ 17 qubits). When we compare VQLS to a direct solver like Gaussian Elimination with $O(N^3)$ complexity we can expect VQLS to show advantage for $N > 750$ ($>$9 qubits).

The primary interest in quantum computing of the SLE is the problems of large dimension. Hence, for a prospective SLE quantum algorithm it is of interest to reach a good performance with increase of N, i.e., the scalability. The scalability of VQLS is limited by the quantum noise and appearance of barren plateaus in the optimization landscape. The latter is a consequence of vanishing gradients, a well-known problem in classical \cite{deepl} and quantum machine learning algorithms \cite{McClean2018}. Recent papers by Mclean et al. \cite{McClean2018} and Cerezo et al. \cite{Cerezo2021} demonstrated the appearance of the barren plateaus in VHQCA.

The quantum noise has been a stumbling block in the quantum computer implementations of all quantum and quantum-classical algorithms. Recent studies \cite{wang2021noiseinduced} have shown that quantum noise can induce barren plateaus. The ansatzes for both local and global cost functions are susceptible to noise-induced barren plateaus in the NISQ era applications, though local cost functions are considered more resistant to the barren plateaus \cite{Cerezo2021}. The noise is dependent on the number of gates, and thus, correlated to the quantum depth of an ansatz circuit \cite{Salas2008}\cite{YulunP}. Hence, a smaller quantum depth will result in a smaller overall damaging effect of the quantum noise. 

In the present work, we address the scalability problem of VQLS by developing an algorithm with evolving, dynamic ansatz as inspired in part by the adaptive VQE \cite{Grimsley2019} and NASNet \cite{zoph2018learning}. The idea of using a variable structure ansatz has been explored for Quantum Machine Learning (QML) \cite{bilkis2021semiagnostic} and for VQE \cite{rattew2020domainagnostic} and was termed Quantum Architecture Search \cite{zhang2020differentiable}\cite{du2020quantum}. Ostaszewski et al. \cite{Ostaszewski2021structure}, demonstrated that variation of the ansatz structure along with the variational parameters in VQE significantly improved the algorithm performance. Layerwise learning of ansatz layers for quantum neural networks was investigated by Skolik et al. \cite{Skolik2021}, while Rattew et al. \cite{rattew2020domainagnostic} developed evolutionary algorithm to grow the VQE ansatz. Our adaptive algorithm does not aim to improve the computational complexity of VQLS. It has the same polylogarithmic complexity as standard VQLS. We aim to improve the performance of VQLS in presence of quantum noise. Our work focuses on varying the ansatz for VQLS by gradual change of the number of layers in ansatz until desired accuracy was reached while keeping used quantum resources at minimum. We evaluate our approach against the standard VQLS algorithm with hardware efficient ansatz \cite{Kandala2017} by comparing the quantum depths of the standard and dynamic VQLS ansatzes in a number of SLE test examples. We define and use in Section \ref{sec:Results} a new metric termed “total resource cost” for a more stringent comparison between the two approaches to VQLS in both, presence and absence of the quantum noise. To induce the quantum noise, we use IBM Qiskit noise models \cite{Qiskit} in the simulators. The motivation behind this study is the reduction of the effective quantum depth in the algorithm, thereby limiting the noise and allowing increase of the SLE dimension $N$ in the VQLS implementations.

The paper is organized as following: Section \ref{sec:Methods} describes the methods used in the paper. In Section \ref{sec:Results} we show and discuss the results, while our conclusions are presented in Section \ref{sec:Conclusion}.

\section{Methods}\label{sec:Methods}

The cost functions and ansatz circuits involved in the standard VQLS method \cite{bravoprieto2020variational} are reviewed in subsections \ref{2a} and \ref{2b}. The new dynamic ansatz, the algorithm for its evolution through iterations, classical optimization as well as relevant quantum noise simulations are presented in subsections \ref{2c}, \ref{2d}, and \ref{2e}. 

\subsection{The cost function and the quantum circuits in VQLS} \label{2a}
 \begin{figure}[b]
  \centering
  \includegraphics[width = \columnwidth]{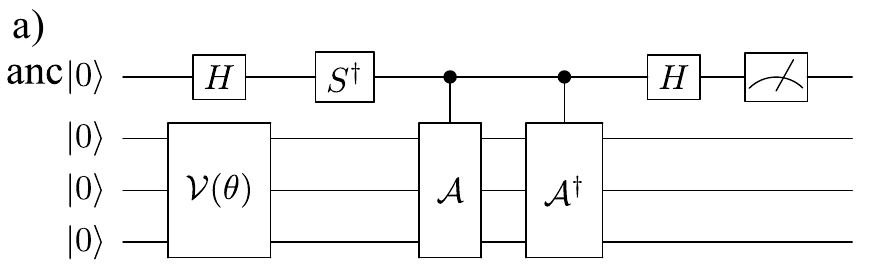}
  \includegraphics[width = \columnwidth]{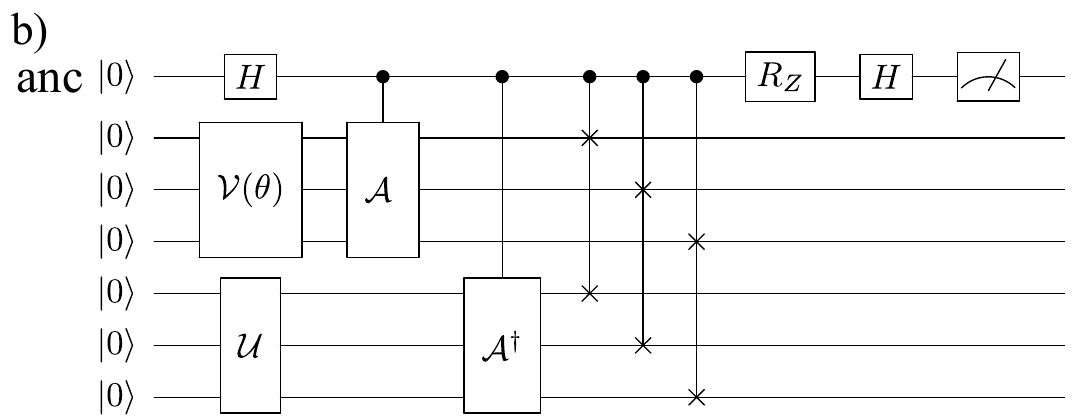}
  \caption{Quantum circuits for VQLS: a) Hadamard test to calculate denominator, and b) swap test to calculate numerator of the cost function for a system of size $N = 8$.}
  \label{fig:hadswap}
 \end{figure}
 The cost function is defined in VQLS \cite{bravoprieto2020variational} as the overlap of states $A\ket{x}$ and $\ket{b} = U\ket{0}$, where $U$ is an unitary operator which is the state initialization circuit as defined in \cite{stateinit}. The trial state is encoded by a variational ansatz $\ket{x} = V(\boldsymbol{\theta})\ket{0}$, with a set of variable parameters, expressed as a vector of rotation angles $\boldsymbol{\theta} = [\theta_0, \theta_1, ..., \theta_m]$. Since matrix $A$ is an arbitrary matrix, it is decomposed into a linear combination of unitary matrices of the form $ A = \sum_l c_l P_l$ where $c_l$ are complex numbers and $P_l$ are Pauli matrices \cite{team_2021}. The decomposition process is a preparatory, one-time, process which we illustrate a few examples in the Supplemental Information (section S1). The Hamiltonian $H_G$ and equation for the global cost function $C_G$ are defined in \cite{bravoprieto2020variational} as
 \begin{equation}
  H_G = A^{\dagger} (1 - \langle b|b \rangle)A
 \end{equation}
 \begin{equation}\label{globalcost}
  C_G = 1 - \frac{|\langle b|\psi \rangle|^2}{\langle \psi|\psi \rangle}
 \end{equation}
with $\ket{\psi} = A\ket{x} = AV(\boldsymbol{\theta})\ket{0}$. Although the local cost functions are more resistant to barren plateaus \cite{Cerezo2021}, we will use the global cost function for which the quantum circuits are simpler than those for the local cost functions. We note that we compare strategies for relatively small ansatzes (up to eight qubits), for which barren plateaus usually do not present a critical problem. 

To calculate the denominator and numerator of the global cost function, Eq. (\ref{globalcost}), one makes use of the Hadamard and swap tests \cite{bravoprieto2020variational}, shown in Fig. \ref{fig:hadswap} on the example of $N=8$. The $S^{\dagger}$ and $R_Z$ gates are used only when measuring the imaginary part of the cost function. The main advantage of both tests is that only the ancilla qubit needs to be measured. While the Hadamard test can also be used to calculate the numerator of the cost function, it leads to a more complex quantum circuit with multiple unitaries needed to be controlled. In contrast, swap test requires twice the number of qubits than the Hadamard test but yields a much simpler circuit with fewer controlled unitaries.

\subsection{Static and Dynamic Ansatzes} \label{2b}
We discern two algorithms, the Algorithm with Static Ansatz (ASA) and with Dynamic Ansatz (ADA). In ASA, the total number of layers “$d$” is predefined and fixed during the VQLS run. In ADA the number of layers evolves, depending on the instantaneous value of the cost function. ADA avoids search in an initially large Hilbert space. Instead, the algorithm starts finding the minima in a small subspace spanned by the single layered ansatz, then another layer of ansatz is appended, and the algorithm moves into a larger subspace. This procedure of adding new layers keeps repeating until value of the cost function reaches the pre-defined convergence criterium.

\begin{figure}[h]
  \centering
  \includegraphics[width = \columnwidth]{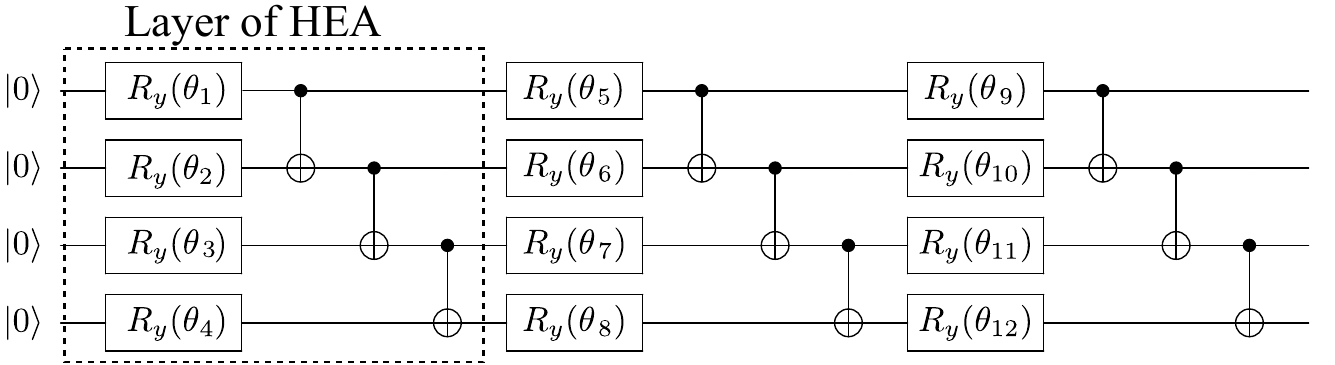}
  \caption{The layered structure of hardware efficient ansatz for SLE with $N = 16$.}
  \label{fig:ansat}
 \end{figure}

Like in ASA \cite{bravoprieto2020variational}, we use in ADA the basic hardware efficient ansatz with layered structure to generate a trial state $\ket{x}$. Each layer consists of a sublayer of rotation gates, followed by a sublayer of entanglers, as illustrated in Fig. \ref{fig:ansat} for $N=16$. This structure is repeated for a total number of “$d$” layers, as needed. The minimum number of layers in ASA estimated to guarantee convergence is given by 
\begin{equation}\label{dmin}
 d_{min} = \frac{2^n}{n}
\end{equation}               
where $n$ is the number of qubits. The reasoning behind the Eq. (\ref{dmin}) is that the vector $\ket{x}$ contains $2^n$ elements, hence at maximum we need one parameter per element. The denominator comes from the fact that hardware efficient ansatz contains $n$ parameters at one layer. The number of layers for ASA is $d_{min}$

The motivation behind ADA is to improve the performance in comparison to the “standard” VQLS with ASA by reducing the total quantum depth of the ansatz, accumulated through iterations and evolution of the number of layers, thereby reducing damaging effects of noise. This is achieved for some types of the SLE systems, as discussed in Section \ref{sec:Results}, thus reducing the quantum noise.
 
\subsection{Classical Optimization} \label{2c}
 In the classical part of the VQLS algorithm we apply the gradient-based optimization utilizing the gradient descent method \cite{ruder2017overview} defined by the equation
 \begin{equation}\label{eqn:graddesc}
 \boldsymbol{\theta_{t+1}} = \boldsymbol{\theta_t} + \delta \nabla_{\boldsymbol{\theta_t}} C_G(\boldsymbol{\theta_t})
\end{equation}  
where $\boldsymbol{\theta_t}$ is the value of the parameter vector at iteration $t$, $\delta$ is the step size of the optimizer and $C_G$ is the cost function parameterized by $\boldsymbol{\theta_t}$. The step size is an input to VQLS algorithm. Choosing too large value of $\delta$ may result in lack of convergence while too small value results in long runtimes of VQLS. Below we explain how the gradients $\nabla \theta_j$ are calculated in Eq. (\ref{eqn:graddesc}). 

In classical computing, backpropagation has been the method of choice for calculating gradients in a neural networks. The ansatz bears some resemblance to the neural network, more specifically to the parameterized connected network of parameters which are “trained” using an optimizer. Still, the ansatz cannot be optimized using backpropagation, since this requires the measurement of intermediate layers of the “neural network”, which would cause collapse of the quantum wave function, disabling its transfer to the next layer. Hence, approaches which do not require intermediate measurements like are finite differences, automatic differentiation, and parameter shift rule are choices for gradients calculations in a quantum ansatz \cite{paramshift}. We use the parameter shift rule which provides exact gradients and is defined by equation,
\begin{multline}
\label{eqn:paramshift}
\frac{\partial C_G(\boldsymbol{\theta})}{\partial \theta_i} = \frac{C_G\left([\theta_1, ..., \theta_i + \frac{\pi}{2}, ..., \theta_k]\right)}{2}\\ - \frac{C_G\left( [\theta_1, ..., \theta_i - \frac{\pi}{2}, ..., \theta_k]\right)}{2}
\end{multline}
where $\boldsymbol{\theta} = [\theta_1, ..., \theta_i, ..., \theta_k]$, where $k$ is the iteration dependent number of parameters and $i = 1, 2, ..., k$.

This equation requires two cost function computations for gradient calculation of each parameter. The cost function evaluations are executed on a quantum computer and hence are affected by quantum noise. The time $t_L$ spent in each optimization loop which calculates all gradients can be is calculated from 
 \begin{equation}\label{eqn:timeopt}
 t_L = 2\nu t_c
\end{equation}
where $\nu$ is the total number of variational parameters and $t_c$ is the time for each cost function computation. In case of the hardware efficient ansatz $\nu$ is obtained as the number of qubits multiplied by the number of layers, which can be conjectured from Fig. \ref{fig:ansat}. 
	
We performed a benchmarking experiment comparing various simulators for time taken by each of them to calculate the cost function as well as for time taken for gradient calculations and parameter updates. The details are presented in Supplemental Information, section S7. From Table S9 it is clear that Pennylane-qulacs is on average about 2 times faster than Pennylane default and about 3 times faster than Pennylane-qiskit in calculations of the gradients. 

\subsection{Evolution of ansatz in ADA} \label{2d}
\begin{figure}
  \centering
  \includegraphics[width = \columnwidth]{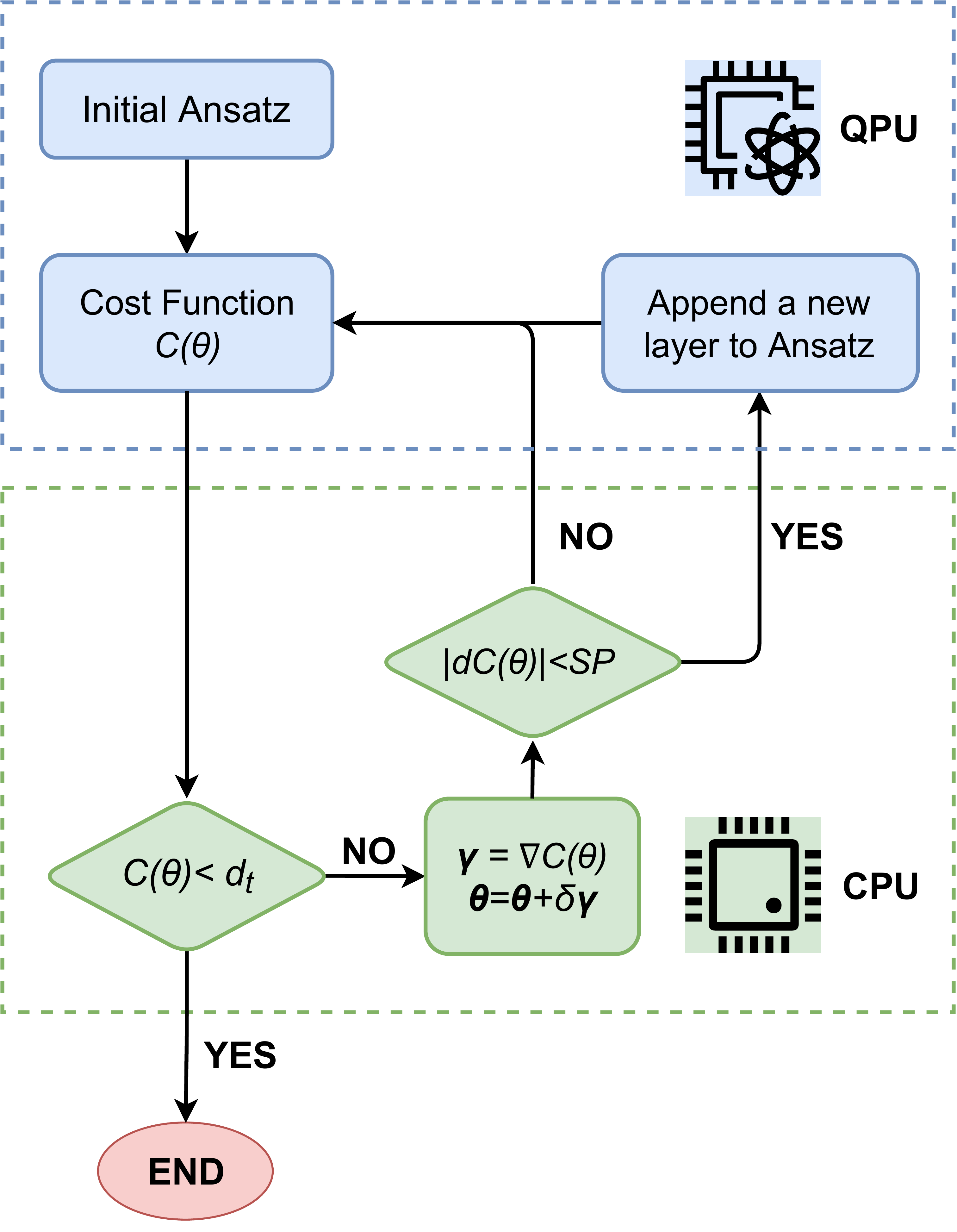}
  \caption{Flowchart for ADA algorithm}
  \label{fig:flowchart}
 \end{figure}

As discussed in subsection \ref{2a} the idea behind ADA is to start with a small configuration space, and increase it successively, limiting the total number of layers when convergence is reached. Initially, we limit the search space to a subspace smaller than the one for a hardware-efficient ansatz with $d_{min}$ layers (Eq. (\ref{dmin})). After finding the minima in the subspace, a check of overall convergence is performed. If convergence is not reached, a new layer of ansatz is appended to the existing ansatz (Fig. \ref{fig:flowchart}). The procedure is continued until the convergence is reached, i.e. value of the cost function goes below a specified threshold $d_t = \frac{\epsilon^2}{\kappa^2}$. To control the speed of appending layers in ADA, we define a hyperparameter which we name the switching parameter (SP). When the difference in costs of two successive optimization steps drops below SP, a new layer is appended. The switching parameter governs the dynamical nature of the ansatz. 

For example, one would have potentially lesser layers at the cost of increasing number of optimization steps, but setting the SP low. Conversely, setting the SP too high would add too many layers before the maximum predefined quantum depth is reached. In this case, the ADA performs like the ASA in terms of both the number of iterations and the Total Resource Cost (to be defined in Section \ref{sec:Results}). 

\begin{figure}
  \centering
  \includegraphics[width = \columnwidth]{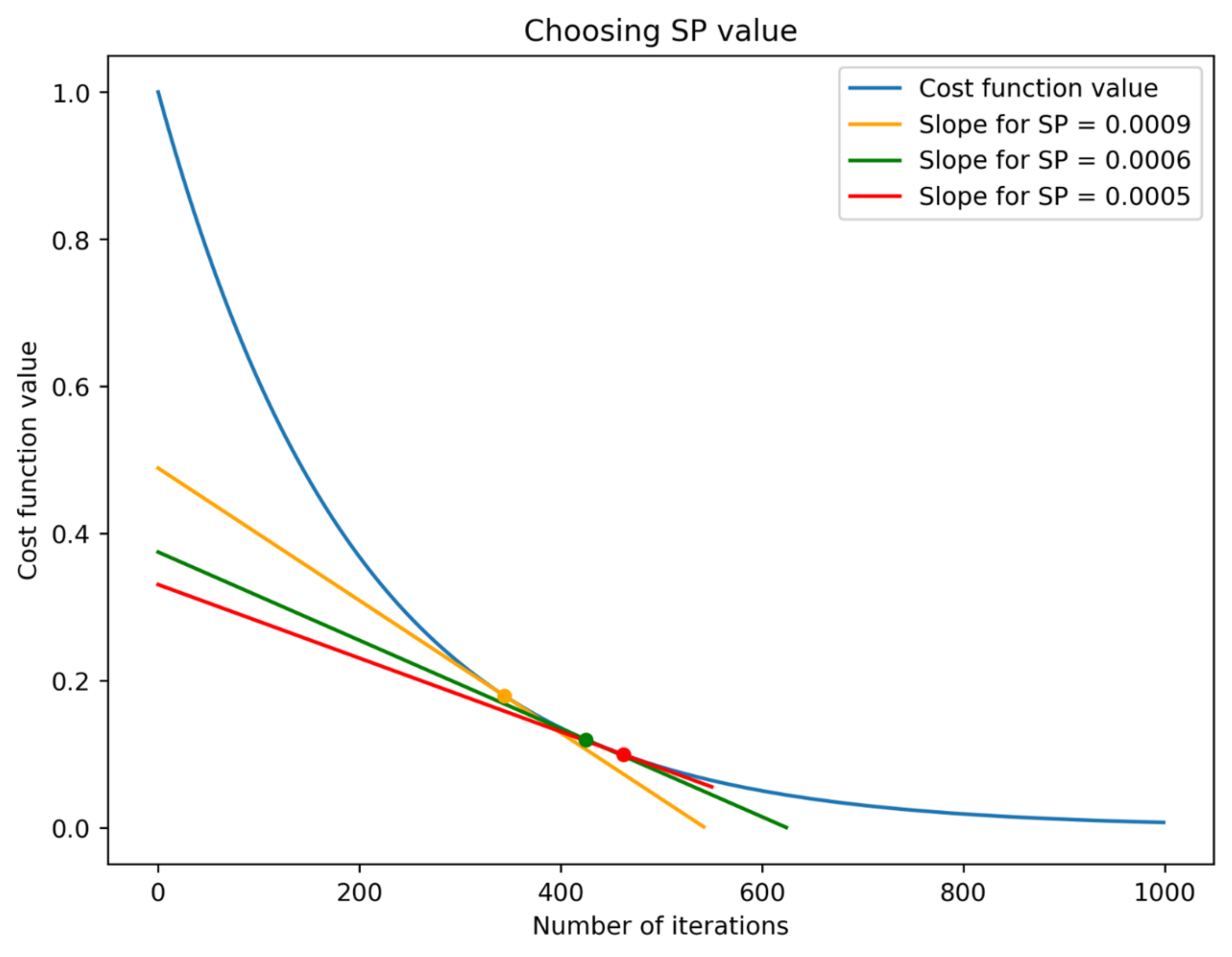}
  \caption{Effect of different SP values on the number of iterations for addition of a new layer}
  \label{fig:SP_graph}
 \end{figure}

A good way to determine value of SP is by following formula,
\begin{equation}\label{eqn:SPchoose}
 SP = \frac{1-d_t}{n_{iterd}}
\end{equation}
where $d_t$ is the stopping threshold (given by the user, which depends on the desired precision ($\epsilon$) of solution $d_t = \frac{\epsilon^2}{\kappa^2}$) and $n_{iterd}$ is the maximum allowed number of iterations which is also set in advance by a user. There is no guarantee that the algorithm will converge by any specified number of iterations. Setting $n_{iterd}$ low will make SP value high, thereby making ADA algorithm to add layers faster. Conversely, setting $n_t$ to a higher value will make SP value low. The idea behind the choice of SP can be understood from the figure \ref{fig:SP_graph}
SP value is compared with the difference of cost function values between two successive optimization iterations to decide the addition of layer of ansatz. Essentially what the algorithm does is  comparison  of the slope of the cost function at each iteration with slope of SP. In figure \ref{fig:SP_graph} the blue line approximates values of cost function against the number of iterations. The other three lines indicate the slopes of different values of SP obtained from the Eq. \ref{eqn:SPchoose}. The points where they are tangent to the blue curve indicate the points where an addition of a new layer is triggered. The $SP = 0.0009$ line will trigger the addition of layer at iteration 344 as compared to $SP = 0.0005$ at iteration 462.
\subsection{Noise Simulation} \label{2e}
We study the effect of noise on VQLS with ADA and ASA using PennyLane \cite{bergholm2020pennylane} with the access to IBM QASM noise simulator. The simulator uses a noise model generated by the calibration data at “ibmq\_16\_melbourne” as an approximation to the noise in the real quantum computer. This noise model is contains depolarizing errors on both single and two-qubit gates, thermal relaxation errors for simulating decoherence, and single-qubit readout errors on all individual measurements, as explained in the Supplemental Information (Section S7). We compare the ASA and ADA in presence of noise in Section \ref{subnoise}.

\section{Results}\label{sec:Results}
All experimental results were obtained using Xanadu's PennyLane library \cite{bergholm2020pennylane}. For the simulations we chose the qulacs backend \cite{suzuki2020qulacs} on basis of the results and discussion in subsection \ref{2c}. For the experiments with noisy simulations we use IBM QASM simulator. The optimization method is utilized with the gradient descent (Eq. (\ref{eqn:graddesc})) with parameter shift rule as in Eq. (\ref{eqn:paramshift}), and with a step size of 0.05. To reduce the computation time, we set convergence threshold $d_t$ of the cost function to 0.1 and the maximum number of iterations to 6400. Only the cases where convergence is reached are discussed in our analysis. Interestingly, the cases when convergence is not achieved are appearing simultaneously for both ASA and ADA. This approach is used in all experiments.

\subsection{Total Resource Cost (TRC)}	
To evaluate and compare efficacies of the ADA and ASA , we define a new performance metric, Total Resource Cost (TRC). From Eq. (\ref{eqn:paramshift}), two cost function evaluations are needed per parameter in each iteration of the optimizer. The number of parameters is determined by the number of rotational gates in the ansatz which is proportional to the number of layers in ansatz. These together define the Effective Quantum Depth (EQD) as quantum depth of the circuit accumulated through the iterations. In ADA the EQD is proportional to the sum of all layers in the evolution of the ansatz. The quantum noise in the algorithm implementation is function of the number of gates, which is proportional the EQD.

TRC is defined as the sum of number of layers of each ansatz over iterations. For ASA, this is simply product of the number of iterations needed for convergence, $z_{ASA}$, and the predefined number of layers in ansatz, $d_{ASA}$,
\begin{equation}\label{eqn:TRCasa}
 TRC_{ASA} = z_{ASA}d_{ASA}
\end{equation}

For ADA the TRC is a cumulative sum of number of layers at each optimization step
\begin{equation}\label{eqn:TRCada}
 TRC_{ADA} = \sum^{z_{ADA}}_{i=0} d_{ADA}(i)
\end{equation}
where $z_{ADA}$ is the total number of iterations of VQLS with ADA and $d_{ADA}(i)$ is the number of layers in the iteration “$i$”. For example, assume that ADA has 10 iterations in total with iteration 1 and 2 having 1 layer, 3 and 4 having 2 layers, 5 to 8 having 3 layers and 9 and 10 having 4 layers in total. The $TRC_{ADA} = 2 \cdot 1 + 2 \cdot 2 + 4 \cdot 3 + 2 \cdot 4$. For ASA, even if we have 7 iterations with 4 layers the $TRC_{ASA} = 28$, i.e. ASA has a higher resource cost in this example. 
 
 The TRC metric is proportional to the EQD of the variational quantum circuit. Each iteration in ASA contains larger or equal number of parameters than ADA (since $d_{ADA}(i) \leq d_{ASA}$ for all i). Fewer number of parameters means lesser number of gradient calculations, thereby implying more efficient computation in each iteration, but also lesser number of (rotational) gates, which means smaller noise. 

Here we compare two ASA examples, which have different number of iterations, $z_{ASA}$, and different number of layers, $d_{ASA}$, showing that neither of these parameters defines unambiguously the algorithm efficacy. In the first example $z^{(1)}_{ASA} = 10$ and $d^{(1)}_{ASA} = 4$. In the second example $z^{(2)}_{ASA} = 20$ and $d^{(2)}_{ASA} = 2$. We also assume that each layer in both cases contains R rotational gates, i.e. R optimization parameters. Since there are two cost function evaluations per parameter, the total number of the cost evaluations is the same for both cases, i.e. $4\cdot10\cdot2=2\cdot20\cdot2=80$, implying the same running time. This is correctly reflected by the equality $TRC^{(1)}= TRC^{(2)}=40$, but not with either number of iterations or number of layers. 

The rationale behind introducing TRC is that, unlike classical computing, quantum computing is heavily affected by quantum noise. The noise is directly proportional to depth of a quantum circuits. Polylogarithmic computational complexity of the quantum algorithm vs. polynomial computational complexity of the classical algorithms does not capture the effect of noise on the algorithm. Even when two algorithms have the same computational complexity, the needed number of operations could be different because of a different quantum circuit depth. Hence, we need a metric that tracks the usage of resources (i.e. number of gates) over all iterations. Since addition of each gate increases noise in the circuit, the TRC measures how the proposed algorithm is noise resilient.
\subsection{Variation of the number of qubits}
In this experiment we use 20 randomly generated SLEs for each of the sizes $N = 16, 32$ and $256$ (i.e. with 4, 5 and 8 qubits), using the PennyLane software library. The condition numbers of SLE are varied from 1 to 20. About 50\% of these tests reach convergence for which we compare TRCs, final number of layers and total number of iterations with both ASA and ADA. The averaged results are shown in Table \ref{tab:Noq}. The detailed results can be seen in tables S2-S6 of Supplemental Information (sections S2- S5). 

%table
\begin{table}
\caption{\label{tab:Noq}Comparison of arithmetically averaged experimental data for ADA and ASA of VQLS.}
\begin{ruledtabular}
\begin{tabular}{ccccccc}
\multicolumn{1}{c}{Number} &\multicolumn{2}{c}{TRC}&\multicolumn{2}{c}{Final Layers}&\multicolumn{2}{c}{Total iterations}\\
 of qubits & ADA & ASA & ADA & ASA & ADA & ASA\\ \hline
 4& 1777.8 & 1949.2 & 3.4 & 4 & 613.3 & 487.3 \\
 5& 2385.68 & 2910.81 & 4.5 & 6 & 572.77 & 485.13\\
 6& 6442.53	& 7366.93 & 6.33 & 8 & 890.06 & 920.86\\
\end{tabular}
\end{ruledtabular}
\end{table}

The arithmetic averages of $TRC_{ADA}$ are smaller (up to ~20\%) than these in $TRC_{ASA}$ for all considered cases. This implies the faster calculations and smaller total noise of ADA. We also compare the number of layers of ansatz in the final iteration, which we term as “final layers”. On average ADA finishes with smaller number of final layers than ASA. As the number of qubits increases, the ADA requires fewer number of iterations than ASA to converge. 

\subsection{Variation of the matrix condition number}
Here we use 25 SLEs of dimension $N=32$ (5 qubits). Condition number is calculated by taking the ratio of the largest to the smallest eigenvalue of matrix A. The number of layers in ansatzes was 6 (refer Eq. (\ref{dmin})) for ASA while the maximum number of layers was set to 6 for ADA. We varied the condition number in range 1 to 20 (see Table \ref{tab:condn}). For each condition number, the experiments at four systems are performed and these all reach convergence. We calculate TRC for both ASA and ADA, and and for comparison we calculate Average of the Relative TRC (ARTRC) deviation:
\begin{equation}\label{eqn:TRCdeviation}
 \frac{\Delta TRC}{TRC} = \frac{TRC_{ADA} - TRC_{ASA}}{TRC_{ADA}}
\end{equation}


\begin{table}[h]
\caption{\label{tab:condn}
Comparison of ARTRC deviation between ASA and ADA results with respect to the condition number for SLEs of a fixed size 32 (5 qubits).}
\begin{ruledtabular}
\begin{tabular}{cc}
\textrm{Condition Number}&
\textrm{ARTRC deviation}\\
\hline
1  & 0.0625 \\
3.684 & 0.41 \\
7.899 & 0.31 \\
13.572 & 0.27 \\
20.651 & -1.18 \\
\end{tabular}
\end{ruledtabular}
\end{table}

The averaged results are presented in Table \ref{tab:condn}, while the details are shown in Supplemental Information (section S3, table S5). The minus sign indicates better performance of ADA.

Better performance is obtained for ASA when the condition numbers are smaller. Up to condition number 5 $TRC_{ADA}$'s are more than 40\% larger. But with further increase of condition number, the performance of the ADA is improving. For poorly conditioned matrices (i.e matrices with a high condition number ~20) the ADA results in 18\% lower TRC than that of ASA.

\subsection{Variation of the matrix sparsity} 
The systems of equations with sparse matrices are common occurrence in science and technology. In our experiments, we define sparsity of a matrix as the ratio of the number of zero matrix elements to the total number of matrix elements. We used matrices of size $N = 16$. The maximum number of layers was set to 4. The sparsity was varied in 4 steps, from 0.9375 to 0.75, while the condition number of the matrices was kept low, between 1 and 1.5. For each sparsity, experiments were performed with ten different matrices. The ARTRC deviation is presented in the table \ref{tab:sparsity} (details in SI, table S6)

\begin{table}[h]
\caption{\label{tab:sparsity}
Comparison of ARTRC deviation between ASA and ADA results with respect to the condition number for systems of a fixed size 32 (5 qubits).}
\begin{ruledtabular}
\begin{tabular}{cc}
\textrm{Sparsity}&
\textrm{ARTRC deviation}\\
\hline
0.9375 & -0.04 \\
0.875 & 0.17 \\
0.8125 & -0.012 \\
0.75 & -0.14 \\
\end{tabular}
\end{ruledtabular}
\end{table}
The sparsity does not significantly impact performance of VQLS with ADA vs ASA, although there is a weak advantage of the ADA results. This advantage is expected to increase for the system matrices with larger condition numbers. 

\subsection{Comparison of ADA and ASA in presence of quantum noise}\label{subnoise}
In the noisy simulations we used systems of dimension $N=16$. More details on how we perform the noise simulations can be found in subsection \ref{2e}.

\begin{table}[h]
\caption{\label{tab:noise}
Comparison of the TRCs for ADA and ASA in presence of noise.}
\begin{ruledtabular}
\begin{tabular}{cccc}
\multicolumn{1}{c}{\textrm{Condition}} &\multicolumn{2}{c}{\textrm{Average TRC}}&\multicolumn{1}{c}{\textrm{\% of cases}}\\
\textrm{Number} & \textrm{ADA} & \textrm{ASA} & \textrm{$TRC_{ADA}<TRC_{ASA}$} \\
\hline
1 &	261.1 & 205.6 & 30 \\
3.6840 & 639 & 625 & 75\\
7.8995 & 542.4 & 550.4 & 80 \\
20.6519 & 699.5 & 1024.5 & 100\\

\end{tabular}
\end{ruledtabular}
\end{table}

From Table \ref{tab:noise}, the TRC of ADA is becoming substantially smaller than of ASA with increase of the condition number and with quantum noise present. Also, in almost 30\% of cases, ADA reached convergence with lesser number of final layers than ASA. In all other cases both algorithms reached the convergence with same number of final layers.
\subsection{Variation of the number of layers in ASA} 
In ASA, the number of layers have to be specified before the algorithm is executed. It is important to have sufficient number of layers so that the ASA is not under-parameterized to reach convergence. However, it is not always easy to predict the number of layers as the input system matrix can vary in terms of condition number and sparsity. Hence, to guarantee convergence with ASA, one might over-parameterize the ansatz by adding more than the needed number of layers. The problem with overparameterization is that the extra layers increase the quantum depth of the circuits used to calculate the cost function, thus increasing the total amount of noise. Due to these factors, setting the right number of layers is essential in ASA. This is not a problem with ADA since it changes the number of layers dynamically without specifying it in advance. 

In the present experiments, the number of qubits is 8. To guarantee convergence, irrespective of the input matrix, the needed number of layers is 32, for which the vector $\ket{x}$ contains 256 elements (more details in Supplemental Information, section S2). With 8 parameters in each layer, 32 layers contain 256 parameters. We purposely under-parameterize the ASA by setting the maximum number of layers to 8 and 16. This is done to show that for some matrices the needed number of layers in ASA is lesser than the $d_{min}$ of Eq. (\ref{dmin}). We utilize VQLS to 21 systems, applying 8 and 16 layers ASA, and also ADA with maximum 8 layers. We found that out of 21 matrices only 6 matrices reached convergence, in each case with $d_t=0.1$. The convergence was reached with the similar values of TRC's for all three case groups. Another interesting observation is that for ASA with 16 layers the algorithm takes fewer iterations than with ansatz with 8 layers but still converges with only slightly higher TRC. The details are shown in SI Section S8, Table S10.
\subsection{Variation of the switching parameter}
\begin{table}[htb]
\caption{\label{tab:sp}
Variation of the TRC and Final number of layers in ADA when SP is varied.}
\begin{ruledtabular}
\begin{tabular}{cccccc}
\multicolumn{1}{c}{} &\multicolumn{4}{c}{\textrm{ADA}}&\multicolumn{1}{c}{\textrm{ASA}}\\
\textrm{SP}& \textrm{$10^{-1}$} & \textrm{$10^{-2}$} & \textrm{$10^{-3}$} & \textrm{$10^{-5}$} & NA \\
\hline
\textrm{Lowest TRC (\%)} & 33 & 14 &	52 & 14 & 0 \\
\textrm{FL\footnote{FL stands for Final Layers} (\%) ADA $< d_{min}$} & 0 &	0 &	66.67 &	71.42 &	NA \\

\end{tabular}
\end{ruledtabular}
\end{table}
In these experiment we use 30 systems of $N=16$ (4 qubits) with varying condition numbers of their matrices A. We consider results for the 21 systems which reached convergence with ADA, with each of the SPs (0.1, 0.01, 0.001, 0.0001) as well as with ASA. Limiting number of layers for ADA is set to 4, which was also the fixed number of layers for ASA. The main objective is to test the effect of the choice of SP on the TRC performance of ADA, which could provide a mean to obtain a good balance between the number of iterations and number of layers in ansatz to reach convergence. 

In Table \ref{tab:sp}, first row shows percentage of cases when ADA with particular value of SP (in columns) has the minimal value of TRC among all 105 cases (4x21 converged with ADA plus 21 converged with ASA). The second row indicates percentage of cases when final number of layers with a particular SP value in columns was less than the $d_{min}$.

As seen in the Table \ref{tab:sp}, some applications of ADA have lesser TRC than ASA. The TRC and final number of layers of ADA and ASA is almost the same when SP is 0.1 (Section S6). The TRC of ADA seems to have a convincing minimum when SP is 0.001. This indicates that there is an optimum value of switching parameter which can be found out empirically from numerical experiments. For example, the optimum value for above system is obviously between 0.00001 and 0.01. From the Table \ref{tab:sp} one can also see that out of all experimental runs, in 67\% cases the final number of layers of ADA is less than that of ASA if SP is 0.001 (Refer SI, section S7 for details).
\subsection{Future directions with ADA}
In this work ADA was used to improve performance of the VQLS algorithm. However, we expect that ADA could be also applicable in VQE and Quantum Machine Learning, since ADA algorithm is agnostic. The future improvements of the algorithm could be done by implementing the popular optimizers like ADAM \cite{DBLP:journals/corr/KingmaB14} and other momentum based gradient optimizers. The momentum value can help to adjust the switching parameter. It would be also interesting to investigate the performance of ADA with non-gradient based optimizers like are Powell and COBYLA \cite{powell}. These have fewer evaluations of the quantum cost function. Another possible improvement can be also made by using optimizers like L-BFGS \cite{fletcher_1987} which use the second order derivatives for faster convergence, but on account of more cost function computations. ADA tends to have fewer parameters than ASA and thus can make use of L-BFGS more feasible for NISQ era. The second order derivative values can be leveraged in ADA for switching decision between the layers.

\section{Conclusion}\label{sec:Conclusion}
Developing an efficient solver for a system of linear equations is one of the central tasks in computing due to its wide use in many physical and engineering applications. Unlike the classical algorithms, VQLS is an algorithm with polylogarithmic scaling with respect to the system size, which promises to offer the quantum advantage even in the NISQ era. The goal of this study was to show that with a good decision making process within a variational algorithm, in particular with VQLS, we can significantly improve its performance in the NISQ era. In our study the decision making process is the variation of the number of layers in the VQLS ansatz, which is the basis for development of our algorithm with dynamic ansatz, ADA. We computationally compared efficiency of ADA with the standard algorithm with static ansatz, ASA, varying the system size (i.e. the number of qubits), condition number and sparsity of the system matrix, as well as switching parameter and number of layers in the hardware efficient ansatz. In presence of noise, ADA tends to outperforms ASA algorithm, quantified by reduction of the total resource cost. Since TRC is directly dependent on the number or layers in the ansatz as well as of the number of iterations, reduction of the TRC results into reduction of the effective quantum depth of the circuit, accumulated through the iterations. ADA also outperformed ASA when the system is ill-conditioned, i.e. when the system matrix has a large condition number. With the increase of number of qubits, ADA starts outperforming ASA not only measured by TRCs but also by number of iterations. While one can always estimate the number of layers for ASA to guarantee convergence, this may often lead to overparameterization, and thus unnecessary increase of the quantum noise. ADA is not susceptible to the overparameterization problem since it doesn’t require specifying number of layers of ansatz in advance. Rather, the algorithm gradually finds the appropriate amount of layers, i.e. the number of optimization parameters, for convergence. Fewer parameters mean fewer cost function evaluations on the quantum computer to calculate gradients. In particular, the values of gradients are less affected by noise. In ADA we define the switching parameter, SP, which controls the addition of new layers in the iteration process. By a proper choice of SP one can adapt the optimum performance of ADA to the noise characteristics of the particular quantum computer. Finally we remark that, ADA is a subset of Quantum Circuit Architecture Search Algorithms. This class of algorithms is useful in optimizing a quantum circuit architecture during the algorithm run. 
\section{Acknowledgments}
HP acknowledges the useful discussions with Dr. Ji Liu. YW was supported by a graduate student scholarship from IACS. All experiments were performed at Seawulf HPC cluster of SBU, and at XSEDE HPCs Comet and Expanse of SDSC.

\bibliography{adavqls}% Produces the bibliography via BibTeX.

\end{document}